\documentclass[aps,prd,eqsecnum,preprint,tightenlines,nofootinbib,showpacs]{revtex4-1}
\usepackage{graphicx,latexsym}

\def\bea{\begin{eqnarray}}
\def\eea{\end{eqnarray}}
\def\be{\begin{equation}}
\def\ee{\end{equation}}

\newcommand{\Pplus}{{\cal P}_+}

\begin{document}

\title{Transitioning from equal-time to light-front quantization \\
in $\phi_2^4$ theory
}
\author{Sophia S. Chabysheva\footnote{Present address: 
Department of Physics, University of Idaho, Moscow ID 83844 USA}}
\author{John R. Hiller$^a$}
\affiliation{Department of Physics and Astronomy\\
University of Minnesota-Duluth \\
Duluth, Minnesota 55812}

\date{\today}

\begin{abstract}

We use the interpolating coordinates studied by Hornbostel to investigate
a transition from equal-time quantization to light-front quantization, in
the context of two-dimensional $\phi^4$ theory.  A consistent treatment
is found to require careful consideration of vacuum bubbles, in
a nonperturbative extension of the analysis by Collins.  Numerical
calculations of the spectrum at fixed box size are shown to yield
results equivalent to those of equal-time quantization, except 
when the interpolating coordinates are pressed toward the light-front
limit.  In that regime, a fixed box size is inconsistent with an
accurate representation of vacuum-bubble contributions and causes
a spurious divergence in the spectrum.  The light-front limit instead
requires the continuum momentum-space limit of infinite box size.
The calculation of the vacuum energy density is then shown
to be independent of the interpolation parameter, which implies
that the light-front limit yields the same spectrum as an
equal-time calculation.  This emphasizes the importance of
zero modes and near-zero modes in a light-front analysis of
any theory with nontrivial vacuum structure.

\end{abstract}


\maketitle

\section{Introduction}
\label{sec:Introduction}

Recently, there has been a resurgence of interest in the spectrum
of two-dimensional $\phi^4$ theory~\cite{Hogervorst,RychkovVitale,%
Rychkov,Pelissetto,DePalma,Bajnok,BCH,Christensen,Katz1,Katz2,Katz3,Katz4,Serone,%
Kadoh,Romatschke},\footnote{For citations of
older work, see \protect\cite{Katz1}.} partly because of what appeared
to be an inconsistency between results from equal-time quantization
and light-front quantization.  Although the apparent inconsistency
has been resolved, as a difference in mass renormalizations~\cite{SineGordon,BCH,Katz1,Katz3},
there remain various issues related to the structure of the vacuum.
In light-front quantization~\cite{BPP,Burkardt,Hiller}, the vacuum is famously 
trivial,\footnote{See, however, the remarks by Collins~\protect\cite{Collins};
Martinovic and Dorokhov~\protect\cite{Martinovic};
and Mannheim, Lowdon, and Brodsky~\protect\cite{MLB}
on nontrivial aspects.} but in equal-time quantization, it is as complex 
as any of the other eigenstates.

More specifically, a direct quantitative check of the difference in
mass renormalizations~\cite{BCH} was not completely successful and
required extrapolation of the light-front mass renormalization from
weaker coupling.  In \cite{Katz3} this failure is 
attributed\footnote{See the discussion in Sec.~2.2 of \protect\cite{Katz3}.}
to an incomplete nonperturbative formulation of the mass renormalization itself.
The mass renormalization relies on a computation of the expectation
value $\langle\phi^2\rangle$ for the square of the field $\phi$, which is
done with a spectral decomposition.  The mass spectrum that was used
in \cite{BCH} lacked the correct behavior near the critical coupling.
This caused the incorrect behavior for the computed value of $\langle\phi^2\rangle$
and created the need for extrapolation.
The incorrect behavior for $\langle\phi^2\rangle$ found in \cite{Katz3}
may also be a numerical artifact that misrepresents the 
underlying theory.  A better understanding requires an
improved treatment of vacuum effects in light-front calculations.

In order to see more clearly what may be happening for the light-front
vacuum, we apply the interpolation procedure championed by
Hornbostel~\cite{Hornbostel}\footnote{There are earlier applications
of interpolation, to two-dimensional QCD~\protect\cite{Frishman},
the Dirac equation~\cite{Ahluwalia}, 
and perturbation theory~\protect\cite{Sawicki},
as well as of quantizations close to the light-cone~\protect\cite{Chen,Elizalde}
applied to two-dimensional QED and QCD~\protect\cite{Franke,Lenz}.}
and emphasized by Ji~\cite{Ji}, in which the (two-dimensional) coordinates are chosen to 
be\footnote{This coordinate transformation is not a Lorentz transformation,
which makes the $c\rightarrow0$ limit technically distinct from the
infinite-momentum-frame limit~\protect\cite{IMF}.}
\be  \label{eq:coordinates}
x^\pm=\frac{1}{\sqrt{2}}[\sqrt{1\pm c}t\pm\sqrt{1\mp c}z],
\ee
with $x^+$ chosen as the time coordinate.  The parameter $c$
ranges from 0 to 1, with 0 being the light-front limit~\cite{Dirac},\footnote{Contrary
to our usual convention but in keeping with an equally common choice, 
$x^\pm$ include a factor of $1/\sqrt{2}$.  This
matches Hornbostel's construction~\protect\cite{Hornbostel}
and simplifies some of the expressions.}
where $x^\pm=(t\pm z)/\sqrt{2}$, and 1 the equal-time limit,
where $x^+=t$ and $x^-=-z$.  The minus sign for the equal-time
spatial coordinate may seem incongruous, but it is a
permissible choice that simplifies the notation.

The conjugate energy and momentum are
\be
p_\pm=\frac{1}{\sqrt{2}}[\sqrt{1\pm c}E\mp\sqrt{1\mp c}p_z].
\ee
Dot products of the momentum and spatial two-vectors are then given by
$p\cdot x=p_+x^+ + p_- x^-$.  The mass-shell condition
becomes
\be
\mu^2=E^2-p_z^2=cp_+^2-cp_-^2+2sp_+p_-,
\ee
with $s\equiv\sqrt{1-c^2}$.  The positive root for $p_+$ yields
\be
p_+=[\sqrt{p_-^2+c\mu^2}-sp_-]/c.
\ee
For the $c=1$ and $c=0$ limits, this expression becomes
\be \label{eq:p-limits}
p_+\rightarrow \left\{ \begin{array}{ll} \sqrt{p_z^2+\mu^2}, & c=1 \\
                            \frac{\mu^2}{2p_-}=\frac{\mu^2}{2p^+}, & c=0, p_->0 \\
                            \frac{\mu}{\sqrt{c}}, & c\rightarrow0, p_-=0 \\
                            \frac{2|p_-|}{c}, & c\rightarrow0, p_-<0. \end{array}\right.
\ee
Clearly, the zero modes ($p_-=0$) and negative $p_-$ states have 
infinite light-front energy and are removed from the spectrum,
as $c\rightarrow0$.

However, these modes can contribute to light-front computations
and, in particular, to vacuum expectation values~\cite{Hornbostel,Ji}.
A standard illustration of this is in the spectrum and VEV of
a free scalar field that has been shifted by a constant.  The
shift introduces to the Lagrangian a term that is linear in
the field; this contributes a term to the Hamiltonian that is
proportional to the spatial average of the field, a constant
that must be built from zero modes absent in ordinary light-front
calculations.  The recovery of the contribution 
can be seen quite clearly in the $c\rightarrow0$
limit, where the Hamiltonian eigenvalue problem in an $x^-$ box 
has an analytic solution for any $c>0$.  A numerical solution 
in a truncated Fock space works just as well.  This is discussed
in Sec.~\ref{sec:shifted}.

For $\phi^4$ theory, the zero-mode contribution is more subtle.
We expand upon the perturbative analysis of Collins~\cite{Collins} to
illustrate how zero-mode contributions to the self-energy
corrections are missed by the standard light-front analysis
yet survive the $c\rightarrow0$ limit.
In Sec.~\ref{sec:phi4}, we will explore what the solutions 
with $c\neq0$ and the $c\rightarrow0$
limit can tell us.  The calculations are done numerically,
in a Fock basis of discrete momentum states in an $x^-$ box.
These lead to a much better understanding of the $c\rightarrow0$ limit.
A fixed box size is shown to be inconsistent with this limit.
Instead, one must consider the continuum limit simultaneously
with the light-front limit.  A summary of these observations
and our conclusions is given in Sec.~\ref{sec:summary}.

\section{Shifted free scalar} \label{sec:shifted}

A free scalar field that is shifted by a constant provides an 
interesting example of the impact of zero modes on a light-front
calculation.  This can be seen explicitly in the $c\rightarrow0$ limit,
where a nonzero contribution is found for the vacuum energy and
the VEV of the field.  These analytic results~\cite{Hornbostel,Ji}
can be replicated in a numerical calculation using a Fock basis of 
zero modes.  We illustrate this here.

The Lagrangian of a free scalar field of mass $\mu$ is
\be
{\cal L}_0=\frac12\partial_\mu\phi\partial^\mu \phi-\frac12\mu^2\phi^2.
\ee
In terms of the interpolating coordinates (\ref{eq:coordinates}),
with arbitrary $c$ and in two dimensions, this becomes~\cite{Hornbostel}
\be
{\cal L}_0=\frac12 c[(\partial_+\phi)^2-(\partial_-\phi)^2]
             +s\partial_+\phi\partial_-\phi-\frac12\mu^2\phi^2.
\ee
The (free) Hamiltonian is
\be
\Pplus^0=\int dx^-(\pi\partial_+\phi-{\cal L}_0),
\ee
with $\pi=c\partial_+\phi+s\partial_-\phi$ and $s=\sqrt{1-c^2}$. The  mode expansion for
the field is
\be  \label{eq:continuousmodeexp}
\phi=\int_{-\infty}^\infty \frac{dp_-}{\sqrt{4\pi w_p}}[a(p_-)e^{-ip\cdot x}
                                  +a^\dagger(p_-)e^{ip\cdot x}],
\ee
with $w_p\equiv\sqrt{p_-^2+c\mu^2}$.  The nonzero commutation relation is
\be
[a(p_-),a^\dagger(p'_-)]=\delta(p_--p'_-).
\ee
The normal-ordered free Hamiltonian can then be written as
\be
\Pplus^0=\int_{-\infty}^\infty dp_- p_+ a^\dagger(p_-)a(p_-)
       =\int_{-\infty}^\infty dp_- \frac{w_p-sp_-}{c}a^\dagger(p_-)a(p_-).
\ee
Similarly, the momentum operator is
\be
{\cal P}_-=\int_{-\infty}^\infty dp_-\,p_-\,a^\dagger(p_-)a(p_-).
\ee

Discretization consistent with discrete light-cone quantization
(DLCQ)~\cite{PauliBrodsky} is invoked
by placing the system in a box $-L<x^-<L$ with periodic boundary
conditions.  The momentum is then discrete, $p_-=n\pi/L$,
as set by the integer $n$; however, unlike DLCQ, $n$ ranges over
all integers, not just the positive ones.\footnote{As shown in the
last line of (\ref{eq:p-limits}),
negative $p_-$ is removed from the spectrum only on the light front, at $c=0$.}
An energy cutoff is
then required for a finite basis, just as for an ordinary 
equal-time calculation.  We do still define a positive
integer $K$ as the resolution~\cite{PauliBrodsky}, so that in
the $c\rightarrow0$ light-front limit, the total momentum is
$P_-=K\pi/L$.  The index $n$ for individual momentum then ranges
from 1 to $K$ in the light-front limit, and momentum fractions 
$p_-/P_-$ are just $n/K$.

The discrete mode expansion for arbitrary $c$ is
\be \label{eq:discretemode}
\phi(x^+=0)=\sum_{n=-\infty}^\infty \frac{1}{\sqrt{4\pi w_n}}
                   [a_ne^{-in\pi x^-/L}+a_n^\dagger e^{in\pi x^-/L}],
\ee
with $w_n\equiv\sqrt{n^2+c\tilde{L}^2}$, 
$[a_n,a^\dagger_m]=\delta_{nm}$,
and $p_-$ replaced by $n\pi/L$.
The free Hamiltonian becomes
\be \label{eq:Pzero}
\Pplus^0=\sum_{n=-\infty}^\infty p_+ a^\dagger_n a_n
=\frac{\mu}{\tilde{L}}\sum_{n=-\infty}^\infty \frac{w_n-sn}{c}a^\dagger_n a_n,
\ee
where $p_+=\frac{\pi}{L}\frac{w_n-sn}{c}$
and $\tilde{L}\equiv\mu L/\pi$.

We now shift the field: $\phi\rightarrow\phi+v$.  The new Lagrangian is
\be
{\cal L}={\cal L}_0-\mu^2 v\phi-\frac12\mu^2 v^2,
\ee
and the Hamiltonian, having dropped a constant, is
\be
\Pplus=\Pplus^0+\Pplus^I,
\ee
with the interaction part
\be
\Pplus^I=\int_{-L}^L dx^- \mu^2 v \phi=\mu\frac{v\sqrt{\tilde{L}\pi}}{c^{1/4}}[a_0+a^\dagger_0].
\ee
In a native light-front calculation, where zero modes are neglected,
this interaction term disappears.
Without this term, the shift in the field and the shift in the energy cannot
be recovered.  However, a calculation for arbitrary $c>0$ succeeds, and the light-front
limit can then be taken.  This was discussed by Hornbostel~\cite{Hornbostel}, and we
repeat the argument here.

The vacuum eigenstate for this case is a coherent state of zero modes
\be
|{\rm vac}\rangle=e^{-\alpha(a^\dagger_0-a_0)}|0\rangle.
\ee
This works because the coherent state is, as always, an eigenstate of
the annihilation operator
\be
a_0|{\rm vac}\rangle=-\alpha|{\rm vac}\rangle
\ee
and, therefore,
\be
\Pplus|{\rm vac}\rangle
=\left[-\frac{\mu w_0}{\tilde{L}c}\alpha a_0^\dagger
+\mu\frac{v\sqrt{\tilde{L}\pi}}{c^{1/4}}a_0^\dagger
-\mu\frac{v\sqrt{\tilde{L}\pi}}{c^{1/4}}\alpha\right]|{\rm vac}\rangle.
\ee
Given $w_0=\tilde{L}\sqrt{c}$, we only need $\alpha=v\sqrt{\tilde{L}\pi\sqrt{c}}$
to eliminate the $a_0^\dagger$ terms and make this coherent state indeed an eigenstate of $\Pplus$,
with an eigenenergy of $-\mu\frac{v\sqrt{\tilde{L}\pi}}{c^{1/4}}\alpha=-\frac12\mu^2 v^2 (2L)$.
This restores the constant originally dropped from the Hamiltonian.
In the light-front limit $c\rightarrow0$, $\alpha$ also becomes zero,
and this state becomes the empty
state $|0\rangle$, but the energy is independent of $c$.  All massive
states are decoupled and remain in the spectrum as free states.

The VEV of the field is given by
\be
\langle {\rm vac}|\phi(0)|{\rm vac}\rangle
=\langle {\rm vac}|\frac{1}{\sqrt{4\pi w_0}}\left[a_0+a_0^\dagger\right]|{\rm vac}\rangle,
\ee
which reduces to 
\be
\frac{1}{\sqrt{4\pi w_0}}(-\alpha-\alpha)
=-\frac{2}{\sqrt{4\pi\tilde{L}\sqrt{c}}}v\sqrt{\tilde{L}\pi\sqrt{c}}=-v.
\ee
This, of course, reflects the original shift in the field.
Obviously, this is independent of the value of $c$.
A non-zero result is obtained because
the vanishing coefficients of zero mode contributions are 
compensated by the $1/c^{1/4}$ divergence in the zero-mode part of the field.

\begin{figure}[ht]
\vspace{0.2in}
\centerline{\includegraphics[width=15cm]{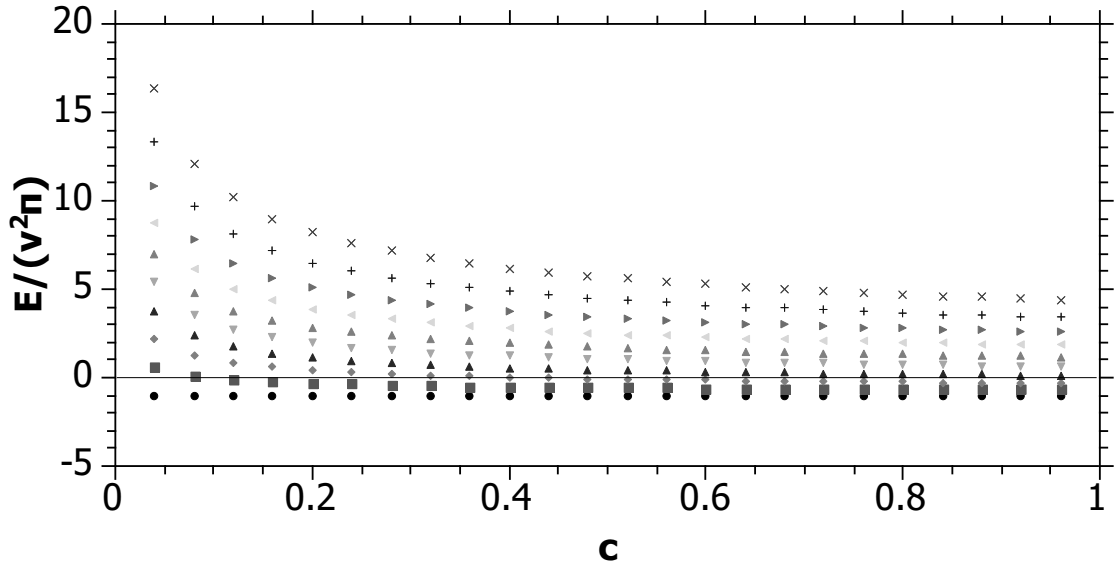}}
\caption{\label{fig:shiftEvsc}
Spectrum for the shifted free scalar in a zero-mode basis,
truncated to an occupation number of 10, as a function
of the interpolating parameter $c$.  Equal-time quantization
corresponds to $c=1$, and light-front quantization to the
limit $c\rightarrow0$.}
\end{figure}

We need not rely on having an analytic solution to see this result
for the vacuum state.  A numerical solution in a finite basis of zero modes
$(a_0^\dagger)^n|0\rangle$, truncated at $n=10$, yields
the spectrum shown in Fig.~\ref{fig:shiftEvsc} as a
function of $c$.  The lowest state's energy is clearly
independent of $c$, with the energies of all higher states
with zero momentum diverging as $c$ approaches zero,
so that zero-mode excitations disappear.

This nontrivial light-front limit provides a connection with
the known results for equal-time quantization.  In the equal-time
approach, the linear interaction term is not lost but makes
a direct contribution to the Hamiltonian.  The solution for
the vacuum state then includes the consequences of the shift
in the field, as can be seen here for $c=1$.  The light-front
limit $c\rightarrow0$ reproduces the results obtained from
equal-time quantization.

A native light-front calculation that does include zero modes
can replicate this result~\cite{Robertson}.\footnote{For 
additional discussion and references, see Secs. 4.1 and 4.2
of \protect\cite{Hiller}.}  The zero-mode part of
the field is determined by a constraint equation derived from
the field equation.  This provides for the correct VEV and 
generates a term in $\Pplus$ that correctly adjusts the
vacuum energy, even though the vacuum state remains the
trivial, empty Fock vacuum.  As shown in \cite{Robertson}, 
the approach can be extended to include $\phi^4$ theory
with spontaneous symmetry breaking, where the mass term
is given the opposite sign and the potential for
the constrained zero mode has definite minima away from zero.
An application to ordinary $\phi^4$ theory~\cite{Pinsky}, in an
attempt to compute the critical coupling, was less successful.
A formulation that solves the constraint equation as an
expansion in the inverse DLCQ resolution has also been 
constructed~\cite{CH}; this provides an alternate approach
for inclusion of zero modes in a DLCQ calculation.

\section{$\phi^4$ theory}  \label{sec:phi4}

To explore these connections between equal-time and light-front
formulations further, we consider two-dimensional 
$\phi^4$ theory, where it is known that equal-time and light-front 
quantizations differ in the vacuum contributions to
mass renormalization~\cite{SineGordon}.  Thus, the remainder
of the paper is an analysis of $\phi^4$ theory in terms
of the interpolating coordinates (\ref{eq:coordinates}).  The 
discrete form of the theory is constructed in the next subsection,
and the results of the numerical solution are discussed for fixed
box size in Sec.~\ref{sec:results}.  These results reveal a
divergence, which is shown to be spurious in Sec.~\ref{sec:bubbles}
by a careful analysis of vacuum bubble contributions.  This
leads to a numerical formulation that avoids the divergence
by varying the box size, as discussed in Sec.~\ref{sec:varied}.

\subsection{Analysis}  \label{sec:analysis}

The Lagrangian for $\phi^4$ theory is
\be
{\cal L}=\frac12(\partial_\mu\phi)^2-\frac12\mu^2\phi^2-\frac{\lambda}{4!}\phi^4.
\ee
We construct the (discrete) interaction Hamiltonian from the $\phi^4$ term as
\be
\Pplus^I=\int_{-L}^L dx^-\frac{\lambda}{4!}:\phi^4:.
\ee
Substitution of the discrete mode expansion (\ref{eq:discretemode}),
with $L=\tilde{L}\pi/\mu$, and evaluation of the now-trivial 
integrals, yields,
\bea \label{eq:PplusI}
\Pplus^I=\mu\frac{g\tilde{L}}{4}\sum_{n_1\ldots n_4}\frac{1}{\sqrt{w_{n_1}\cdots w_{n_4}}}\left[
   \frac{1}{12}(a_{n_1}\cdots a_{n_4}+a^\dagger_{n_1}\cdots a^\dagger_{n_4})\delta_{n_1+\cdots+n_4,0}\right.&& \\
+\frac13(a^\dagger_{n_1}a_{n_2}a_{n_3}a_{n_4}+ a^\dagger_{n_2}a^\dagger_{n_3}a^\dagger_{n_4}a_{n_1})\delta_{n_1,n_2+n_3+n_4} &&
  \nonumber \\
\left.+\frac12 a^\dagger_{n_1}a^\dagger_{n_2}a_{n_3}a_{n_4}\delta_{n_1+n_2,n_3+n_4}\right], &&  \nonumber
\eea
with $g\equiv \lambda/(4\pi\mu^2)$ the dimensionless coupling.

The Hamiltonian eigenstates are constructed as Fock-state expansions
\be
|\psi\rangle=\sum_k \sum_{n_1\cdots n_k}\psi_k(n_1\cdots n_k)
               \frac{1}{\sqrt{k!}}\prod_{i=1}^k a_{n_i}^\dagger|0\rangle.
\ee
To take into account the symmetrization of states with $k$ identical bosons,
we rewrite this sum as
\be
|\psi\rangle=\sum_k \sum_{n_1\geq n_2 \cdots \geq n_k}
              \frac{1}{\sqrt{N_{n_1}!\cdots N_{n_k}!}}\phi_k(n_1\cdots n_k)
               \prod_{i=1}^k a_{n_i}^\dagger|0\rangle,
\ee
where $N_{n_i}$ is the number of bosons with momentum index $n_i$
and the wave functions are related by 
\be
\phi_k=\sqrt{\frac{k!}{N_{n_1}!\cdots N_{n_k}!}}\psi_k.
\ee
The normalization is
\be
1=\langle\psi|\psi\rangle=\sum_k\sum_{n_1\cdots n_k}|\psi_k|^2
      =\sum_k \sum_{n_1\geq n_2 \cdots \geq n_k} |\phi_k|^2.
\ee
The probability $P_k$ for the Fock sector with $k$ bosons is then
given by
\be
P_k=\sum_{n_1\cdots n_k}|\psi_k|^2=\sum_{n_1\geq n_2 \cdots \geq n_k} |\phi_k|^2.
\ee

The eigenstates must satisfy $(\Pplus^0+\Pplus^I)|\psi\rangle=E|\psi\rangle$.
For simplicity, we look for eigenstates at rest, with total $P_-=0$, and either an odd 
or even number of constituents; the Hamiltonian changes particle number by only even amounts
and therefore does not mix odd and even Fock states.  The sums over the number of 
constituents $k$ are then limited to even or odd values.  In particular, we have 
expansions of the form
\be  \label{eq:even-odd}
|{\rm even}\rangle=\psi_0|0\rangle+\sum_n \psi_2(n)\frac{1}{\sqrt{2}}a_n^\dagger a_{-n}^\dagger|0\rangle+\cdots
\ee
\be
|{\rm odd}\rangle=\psi_1 a^\dagger_0|0\rangle+\sum_{n_1,n_2}
      \psi_3(n_1,n_2)\frac{1}{\sqrt{6}}a^\dagger_{n_1}a^\dagger_{n_2}a^\dagger_{-n_1-n_2}|0\rangle+\cdots
\ee
In solving the eigenvalue problem for $\Pplus$, we obtain the spectrum as well as 
the associated Fock-state wave functions $\psi_n$, though we do not display the 
wave functions here.

For the purpose of having a finite numerical matrix calculation,
the infinite Fock basis is truncated both in the sum over constituents
and in energy.  First, the number of
constituents is limited to a maximum of $K$, so that the sum over $k$ in $|\psi\rangle$
is finite.  Second, the total energy of each Fock state, as specified
by the free Hamiltonian (\ref{eq:Pzero}), is limited
to be no more than a fixed energy, $E_{\rm max}$.  

The total energy of a Fock state is given by $\frac{\mu}{\tilde{L}}\sum_n\frac{w_n-sn}{c}$,
where the sum extends over all bosons in the Fock state.  For small $c$,
the individual contributions behave as in (\ref{eq:p-limits}):
\be
p_+=\frac{\mu}{\tilde{L}}\frac{w_n-sn}{c}\rightarrow\left\{\begin{array}{ll} \frac{\mu\tilde{L}}{2n}, & n>0 \\
                                                    \frac{\mu}{\sqrt{c}}, & n=0 \\
                                                    \frac{\mu}{\tilde{L}}\frac{2|n|}{c}, & n<0.
                                  \end{array}\right.
\ee
Thus, for $n\leq0$, the contributions diverge and Fock states with
such constituent momenta will be removed by the energy cutoff as $c$
goes to zero.  For eigenstates with total $P_-=0$, where the integers 
$n$ must sum to zero, there must be at least one constituent with $n\leq0$.  
For such a state, a sufficiently small value of $c$ will cause the energy
cutoff to remove all the Fock states, except the trivial empty state $|0\rangle$.
However, this would be inconsistent with the analysis of the shifted
free scalar, where the addition of a $c$-dependent energy cutoff would have
removed the (infinite set of) Fock states needed to construct the coherent 
state for the vacuum eigenstate.

So, we instead keep the Fock basis unchanged as $c$ changes by imposing 
the energy cutoff at $c=1$ and then leaving the basis fixed when $c$
is decreasing.  Therefore, for all $c$ values, the energy limit on
Fock states is given by 
\be
\frac{\mu}{\tilde{L}}\sum_n \sqrt{n^2+\tilde{L}^2}\leq E_{\rm max}.
\ee

In the following subsection, we pursue a qualitative understanding of the light-front
limit as the parameter $c$ goes to zero. We do not study the dependence on the truncations, 
nor on the box size, in any systematic way.  In equal-time quantization
there has been considerable work by Rychkov and collaborators~\cite{Rychkov}
on the renormalization necessary to reduce the cutoff dependence
and facilitate very accurate calculations with minimal basis sizes.
Attempting this for arbitrary $c$ is certainly of some interest but is
beyond the scope of the present work.

\subsection{Results for fixed box size}  \label{sec:results}

\begin{figure}[ht]
\vspace{0.2in}
\centerline{\includegraphics[width=15cm]{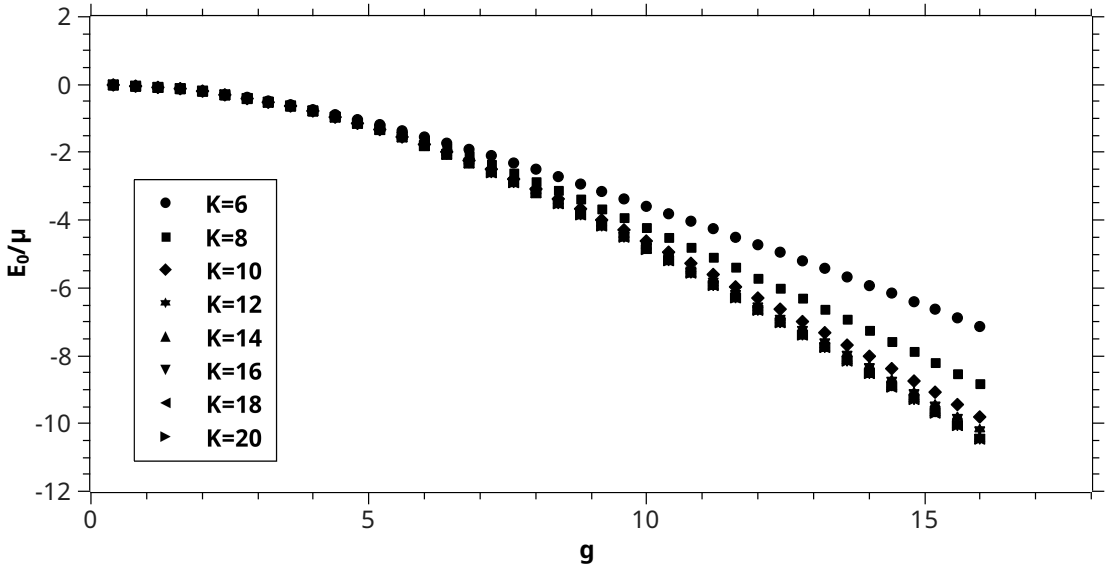}}
\caption{\label{fig:E0vsg}
Even vacuum energy in equal-time quantization as a function of coupling $g$ for a Fock-state energy
cutoff of $E_{\rm max}=20\mu$.  The box size is set by $\tilde{L}=1$.
The maximum number of constituents $K$ is varied up to 20.
}
\end{figure}

\begin{figure}[ht]
\vspace{0.2in}
\centerline{\includegraphics[width=15cm]{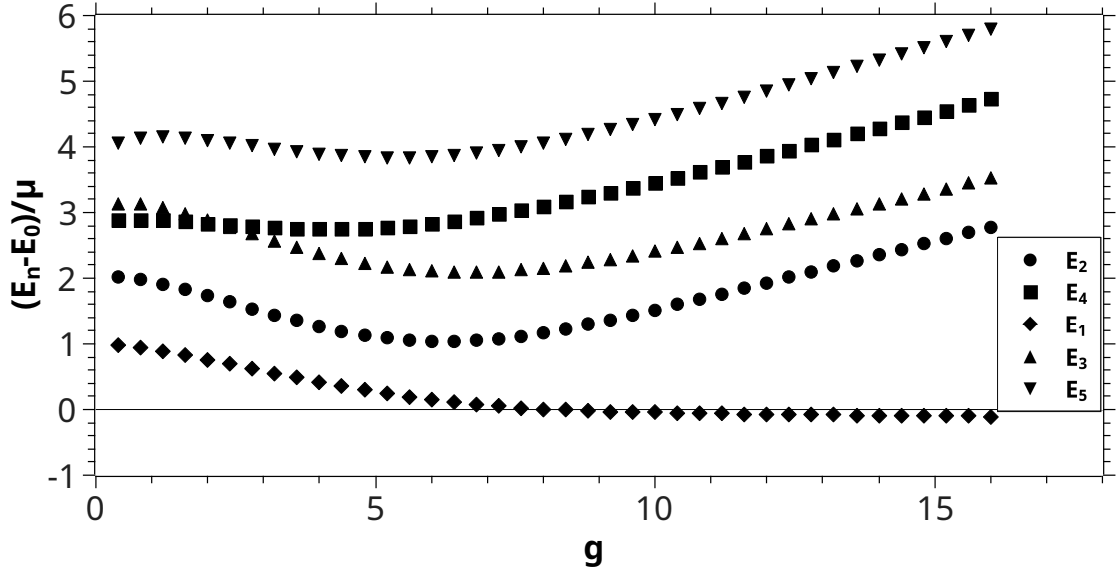}}
\caption{\label{fig:Diffvsg}
Subtracted equal-time spectrum $E_n-E_0$ computed with $c=1$, $E_{\rm max}=20\mu$, 
$\tilde{L}=1$, and up to 20 constituents.  Here $E_n$ is the energy of the
nth level, with $n$ even (odd) for the even (odd) sector.
}
\end{figure}

As a check on the calculation,
the even vacuum energy for equal-time quantization ($c=1$) 
is plotted in Fig.~\ref{fig:E0vsg} as a function of the coupling $g$.
This is computed by solving the eigenvalue problem for $\Pplus$, with
the even eigenstate constructed as in (\ref{eq:even-odd}).
The results for the ground-state (vacuum) energy $E_0$
are equivalent to those of Rychkov and Vitale (RV)~\cite{RychkovVitale},
where $g=6 g_{RV}/\pi$ and $\tilde{L}=L_{RV}/(2\pi)$.
We also plot the subtracted spectrum for equal-time quantization
in Fig.~\ref{fig:Diffvsg}, where the energy $E_0$ of the even vacuum state
is subtracted from the energy of all other states, to show the
energies of physical states above the vacuum. The odd eigenstates
are constructed as in (\ref{eq:even-odd}).  Again, the results are 
equivalent to RV.  In particular, the lowest odd state becomes degenerate
with the even vacuum state at and beyond the critical value of the coupling.

\begin{figure}[ht]
\vspace{0.2in}
\centerline{\includegraphics[width=15cm]{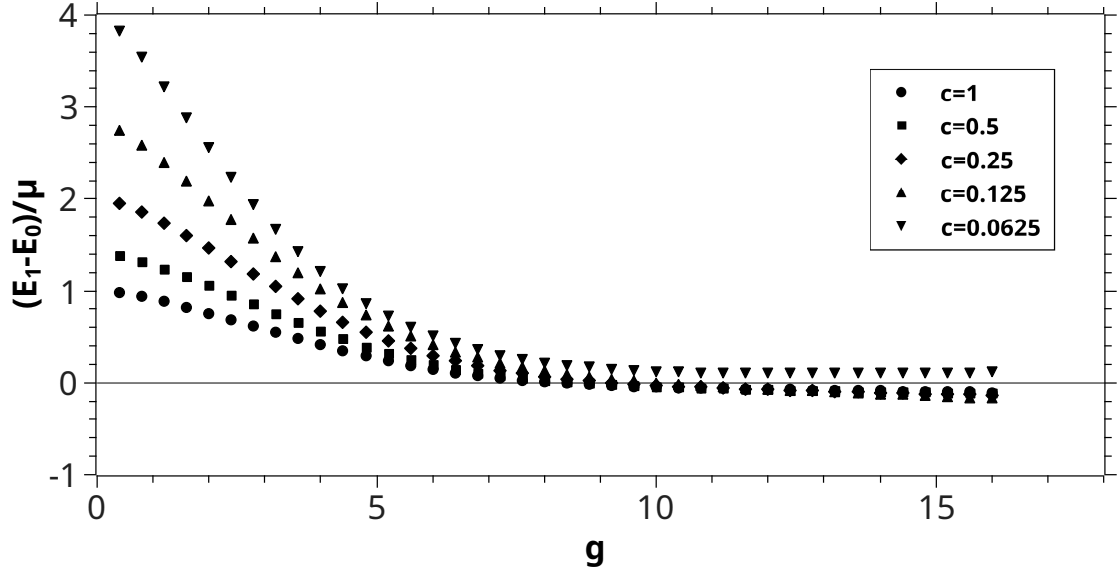}}
\caption{\label{fig:Diffvsgc}
Difference between even and odd vacuum states for 
decreasing values of $c$ and fixed box size $\tilde{L}=1$. 
The difference is larger for smaller $c$;
at $g=0$, the difference is just $\mu/\sqrt{c}$.
}
\end{figure}

With the equal-time results established, we next consider the
variation with $c$, approaching the light-front limit at $c=0$.
Figure~\ref{fig:Diffvsgc} shows how the difference 
between the even and odd vacuum states varies with $g$
for various values of $c$.  For weak coupling the difference increases as
$c$ approaches zero; however, the critical coupling,
where the difference becomes zero, remains essentially the same.
Thus, the $0<c<1$ results are at least qualitatively 
consistent with equal-time quantization, despite that
fact that light-front quantization ($c=0$) is known to give
a different result for the critical coupling~\cite{BCH,Katz1,Katz3}.

\begin{figure}[ht]
\vspace{0.2in}
\centerline{\includegraphics[width=15cm]{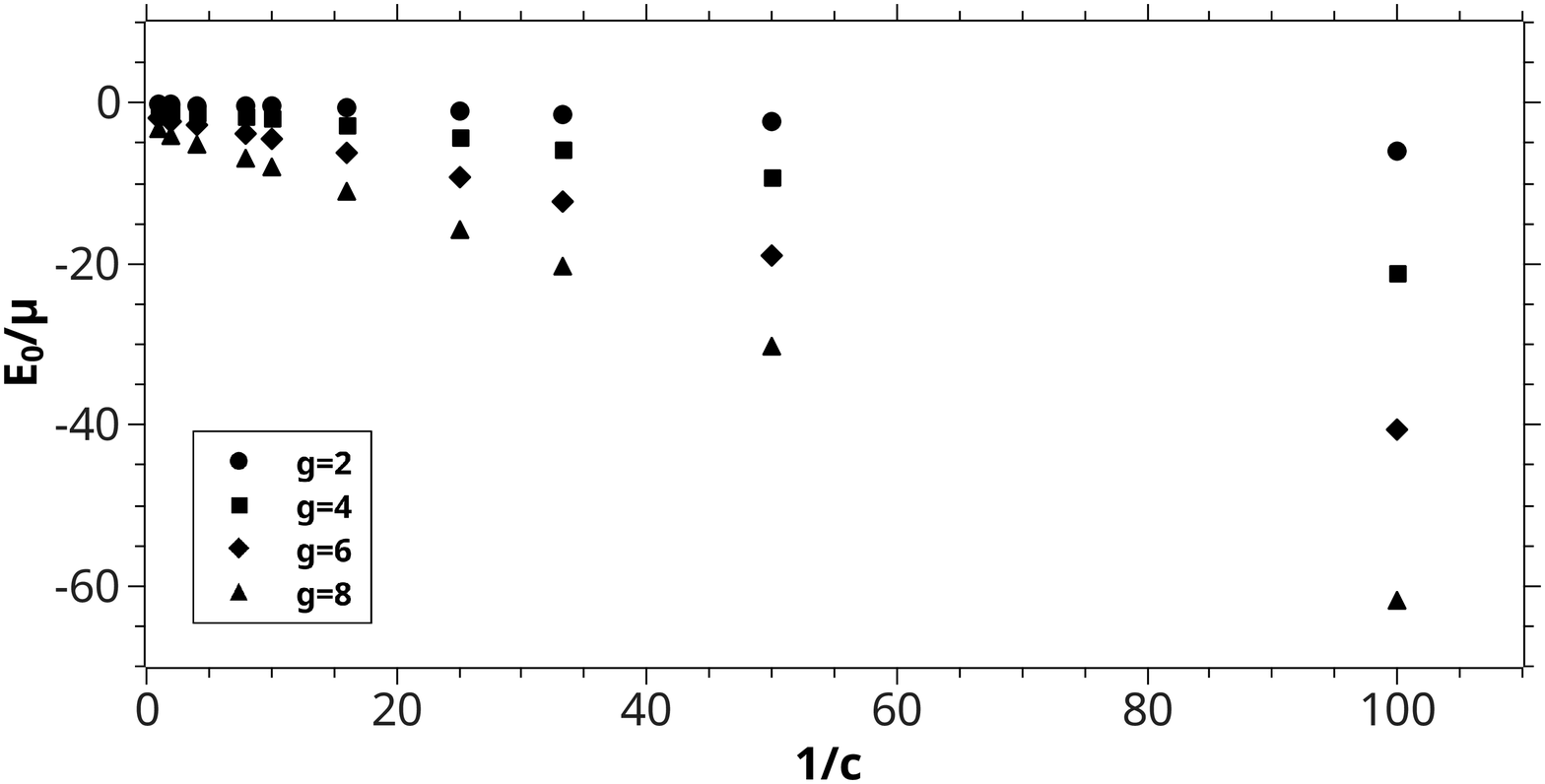}}
\caption{\label{fig:Evsc}
Energy of the even vacuum state as a function of $1/c$ for 
different couplings $g$ and fixed box size $\tilde{L}=1$.  
For fixed $c$, the energy is more negative for larger $g$.
}
\end{figure}

\begin{figure}[ht]
\vspace{0.2in}
\centerline{\includegraphics[width=5cm]{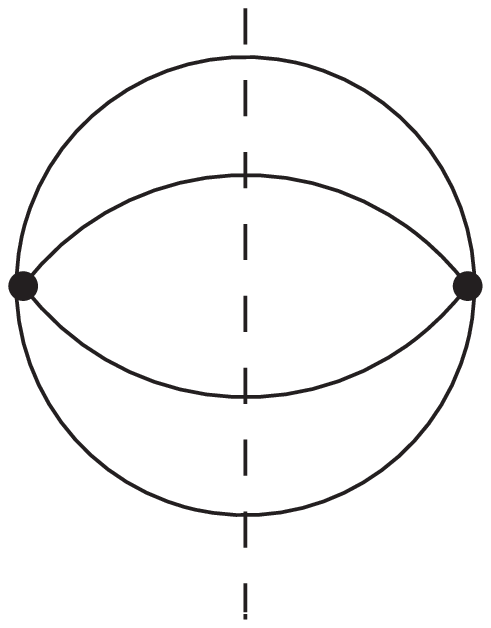}}
\caption{\label{fig:basketball}
Lowest order contribution to the vacuum energy.
}
\end{figure}

To investigate the distinction between the $c\rightarrow0$
limit and a $c=0$ computation, we plot the even vacuum
state energy as a function of $1/c$ for various values
of the coupling $g$ in Fig.~\ref{fig:Evsc}.
As can be seen in the figure, the spectrum appears to diverge as $c\rightarrow0$.
This can be understood~\cite{Polchinski} by considering the simplest contribution to
the vacuum energy, from the `basketball' graph in Fig.~\ref{fig:basketball}.
The zero-mode contribution of this graph to the vacuum energy $E_0$ is
second order in perturbation theory and can be expressed as
\be
\Delta E\sim \frac{g}{w_0^2}\frac{1}{E_0-4 w_0/c}\frac{g}{w_0^2},
\ee
where the middle fraction contains the energy denominator
for an intermediate state with four zero-mode bosons
and is sandwiched between transition matrix elements for 
production and annihilation of four zero modes from and to
the vacuum.  The transition potential is just the first 
line of $\Pplus^I$ in (\ref{eq:PplusI}), which determines
the transitions from the vacuum to four bosons and back to the
vacuum.  The dimensionless individual zero-mode energy is
$w_0=\tilde{L}\sqrt{c}$.  The shift then diverges as
$c^{-3/2}$, which is consistent with the $c$ dependence shown
in Fig.~\ref{fig:Evsc}.  

Clearly, the contributions of vacuum bubbles, such as the contribution
represented by the `basketball' graph in Fig.~\ref{fig:basketball},
require closer inspection, in order to fully understand the calculation.
Recent perturbative analyses~\cite{Collins,Martinovic} of such graphs,
comparing light-front and equal-time calculations, also show that some
care is required.  In fact, the $c\rightarrow0$ transition can help
elucidate the connection between the two quantizations.

\subsection{Vacuum bubbles}
\label{sec:bubbles}

\begin{figure}[ht]
\vspace{0.2in}
\centerline{\includegraphics[width=8cm]{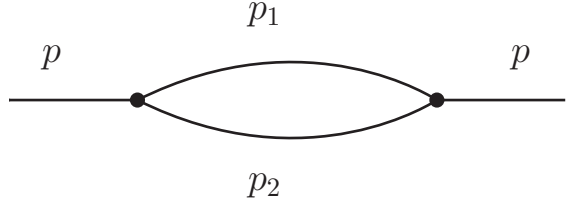}}
\caption{\label{fig:1loopbubble}
One-loop covariant self-energy graph in $\phi^3$ theory.  When $p$ becomes zero,
this is essentially a one-loop vacuum bubble.
}
\end{figure}

Following Collins~\cite{Collins}, we first consider the one-loop
self-energy in $\phi^3$ theory; the graph is given in Fig.~\ref{fig:1loopbubble}.
The invariant function $\Pi(p^2)$ is given by
\be
\Pi(p^2)=-\frac{1}{8\pi^2}\int\frac{d^2k}{[k^2-\mu^2+i\epsilon][(p-k)^2-\mu^2+i\epsilon]}.
\ee
The one-loop bubble is obtained in the $p^2\rightarrow0$ limit, with a
value of $\Pi(0)=-i/8\pi\mu^2$.

\begin{figure}[ht]
\vspace{0.2in}
\begin{tabular}{cc}
\includegraphics[width=5cm]{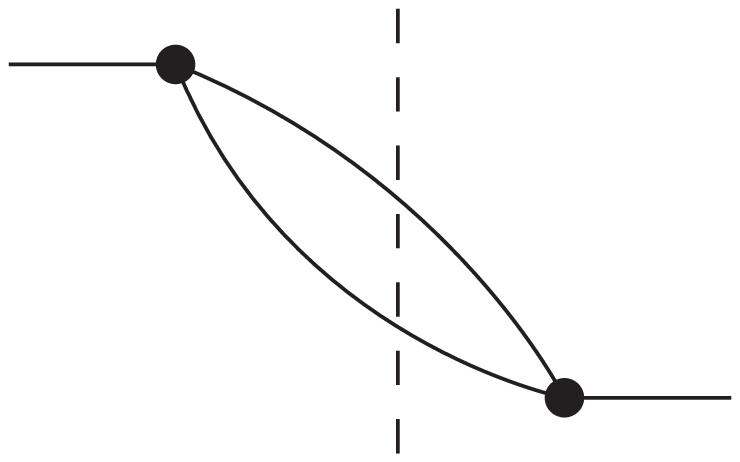} &
\includegraphics[width=3cm]{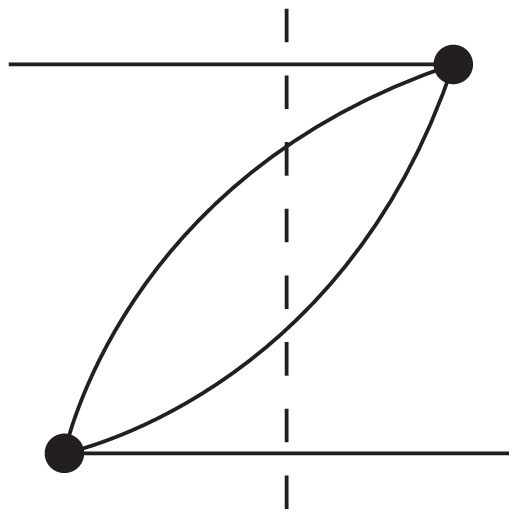} \\
(a) & (b)
\end{tabular}
\caption{\label{fig:1looptimeordered}
Time-ordered graphs corresponding to the covariant graph in Fig.~\protect\ref{fig:1loopbubble}.
}
\end{figure}

The covariant graph is, of course, equivalent to two time-ordered
graphs, shown in Fig.~\ref{fig:1looptimeordered}.  We then have
$\Pi=\Pi_a+\Pi_b$ with
\bea
\Pi_a(p^2)&=&\frac{i}{16\pi}\int \frac{dp_{1-}dp_{2-}}{w_{p_1}w_{p_2}}
           \frac{\delta(p_--p_{1-}-p_{2-})}{p_+-\sum_i^2(w_{p_i}-sp_i)/c}, \\
\Pi_b(p^2)&=&\frac{i}{16\pi}\int \frac{dp_{1-}dp_{2-}}{w_{p_1}w_{p_2}}
           \frac{\delta(p_-+p_{1-}+p_{2-})}{p_+-2p_+-\sum_i^2(w_{p_i}-sp_i)/c}.
\eea
These are perturbative quantities, but a nonperturbative calculation in
$\phi^3$ theory, with the Fock space limited to no more than four constituents,
will yield an eigenvalue condition for the light-front energy $P_+$ that takes the form
\bea \label{eq:P+phi3}
P_+&=&\frac{w_P-sP_-}{c}+\frac{\lambda^2}{32\pi}\int\frac{dp_{1-}dp_{2-}}{w_{p_1}w_{p_2}w_P}
         \frac{\delta(P_--p_{1-}-p_{2-})}{P_+-\sum_i^2\frac{w_{p_i}-sp_{i-}}{c}} \\
 &&+\frac{\lambda^2}{32\pi}\int\frac{dp_{1-}dp_{2-}}{w_{p_1}w_{p_2}w_P}
         \frac{\delta(P_-+p_{1-}+p_{2-})}{P_+-2\frac{w_P-sP_-}{c}-\sum_i^2\frac{w_{p_i}-sp_{i-}}{c}},
         \nonumber
\eea
with an (infinite) two-loop bubble removed.  Details are given in 
the Appendix.  The two self-energy terms in (\ref{eq:P+phi3})
correspond to the two time-ordered perturbative graphs.

In the light-front limit $c\rightarrow0$, $\Pi_b$ makes no contribution,
because vacuum vertices do not exist in the light-front limit, and $\Pi_a$
becomes, with $p_{2-}=p_--p_{1-}$, $w_{p_i}=p_{i-}$, and $x\equiv p_{1-}/p_-$,
\be
\Pi_a(p^2)_{c=0}=\frac{i}{16\pi}\int_0^1 \frac{dx}{x(1-x)}\frac{1}{p^2-\frac{\mu^2}{2x}-\frac{\mu^2}{2(1-x)}}.
\ee
For $p^2=0$, this immediately yields
\be
\Pi_a(0)_{c=0}=-\frac{i}{8\pi}\int_0^1 \frac{dx}{\mu^2(1-x)+\mu^2x}=-\frac{i}{8\pi\mu^2}.
\ee
If instead we take $p^2=0$ first, we have, with $z\equiv p_{1-}/\sqrt{c}\mu$, equal contributions
from each time-ordered graph
\be
\Pi_a(0)=\Pi_b(0)=-\frac{i}{16\pi}\int_{-\infty}^\infty \frac{c\,dp_{1-}}{(p_{1-}^2+c\mu^2)^{3/2}}
        =-\frac{i}{16\pi\mu^2}\int_{-\infty}^\infty \frac{dz}{(1+z^2)^{3/2}}=-\frac{i}{16\pi\mu^2},
\ee
a result independent of $c$. The sum of the two then replicates the full value 
of $-i/8\pi\mu^2$.

It is the latter result that is most important for our calculations, because we calculate
with finite $c$ and take the zero limit last. Although the
analytic integral is independent of $c$, the numerical approximation associated
with taking periodic boundary conditions in a fixed box size is not independent.
A small box forces the momentum-space quadrature points to be widely spaced.
As $c$ approaches zero, the integrand becomes more sharply peaked in $p_{1-}$,
and the important range of integration is not sampled enough, unless the box
is kept sufficiently large.  Thus, as $c$ is decreased, the box size must
be increased such that $c\tilde{L}^2$ is at least of order 1, and the $c\rightarrow0$
limit does not recover DLCQ, which requires a fixed light-front box size.  Instead, the
$c\rightarrow0$ limit must be associated with the continuum limit, where the box
is removed.

To illustrate the convergence that can be obtained by taking the 
continuum limit of $\tilde{L}\rightarrow\infty$, we consider
the numerical estimate of the rescaled quantity
\be
\widetilde{\Pi}_a(0)\equiv 16\pi i \mu^2\Pi_a(0)=\int_{-\infty}^\infty \frac{c\mu^2\, dp_{1-}}{(w_{p_{1-}})^{3/2}}
\simeq \frac12 c\tilde{L}^2\sum_{n=-N}^N \frac{1}{(n^2+c\tilde{L}^2)^{3/2}},
\ee
which has the nominal value of 1.  However, for the numerical calculation
we work at an energy cutoff that determines the range $N$ of the finite sum
over the $p_-$ index $n$.
For the intermediate Fock state of the one-loop bubble where two
constituents have the most energy, the momentum indices are $N$ and $-N$.  
The energy limit is then expressed as
\be
E_{\rm max}\geq\frac{\mu}{\tilde{L}}\sum_n\frac{\sqrt{n^2+c\tilde{L}^2}}{c}
     \rightarrow 2\frac{\mu}{\tilde{L}}\frac{\sqrt{N^2+c\tilde{L}^2}}{c}.
\ee
This leaves
\be
N\leq\sqrt{c}\tilde{L}\sqrt{\left(\frac{\sqrt{c}E_{\rm max}}{2\mu}\right)^2-1}.
\ee
Table~\ref{tab:bubble} shows results for selected values of $c$ and
ranges of box size $\tilde{L}$, with $E_{\rm max}$ varied as needed to
keep the same number of quadrature points for different values
of $c$.  The cutoff in energy does shift the value away from the
nominal value of 1.

\begin{table}[ht]
\caption{\label{tab:bubble}
One-loop and three-loop vacuum bubbles as functions of the dimensionless box size $\tilde{L}\equiv\mu L/\pi$.
The energy cutoff $E_{\rm max}$ and particular values of $\tilde{L}$ are chosen to make the calculation
equivalent for each value of the coordinate parameter $c$.  The values of $N$ indicate the range of
discrete steps in momentum, as determined by the energy cutoff.  In both cases, the value of
$\widetilde{\Pi}_a(0)$ is independent of $c$ and converges in the continuum limit of 
$\tilde{L}\rightarrow\infty$.
}
\begin{center}
\begin{tabular}{r|rrrrr|lr|lr}
\hline \hline
$c$ &   1.0000 &   0.5000 &   0.2500 &   0.1250 &   0.0625 & \multicolumn{2}{c|}{one-loop} & \multicolumn{2}{c}{three-loop} \\
$E_{\rm max}/\mu$ & 20.0000 &  28.2843 &  40.0000 &  56.5685 &  80.0000 &\hspace{0.1in} $N$\hspace{0.1in} & $\widetilde{\Pi}_a(0)$ &\hspace{0.1in} $N$\hspace{0.1in} & $\widetilde{\Pi}_a(0)$ \\
\hline
$\tilde{L}$ & 1.0000 &   1.4142 &   2.0000 &   2.8284 &   4.0000 &\hspace{0.1in}  9 & 1.00696 &\hspace{0.1in}  9 & 7.03843 \\
$\cdot$     & 2.0000 &   2.8284 &   4.0000 &   5.6569 &   8.0000 &\hspace{0.1in} 19 & 0.99482 &\hspace{0.1in} 19 & 7.02181\\
$\cdot$     & 3.0000 &   4.2426 &   6.0000 &   8.4853 &  12.0000 &\hspace{0.1in} 29 & 0.99487 &\hspace{0.1in} 29 & 7.01538\\
$\cdot$     & 4.0000 &   5.6569 &   8.0000 &  11.3137 &  16.0000 &\hspace{0.1in} 39 & 0.99491 &\hspace{0.1in} 38 & 7.01133\\
$\cdot$     & 5.0000 &   7.0711 &  10.0000 &  14.1421 &  20.0000 &\hspace{0.1in} 49 & 0.99494 &\hspace{0.1in} 48 & 7.01264\\
$\cdot$     & 6.0000 &   8.4853 &  12.0000 &  16.9706 &  24.0000 &\hspace{0.1in} 59 & 0.99495 &\hspace{0.1in} 58 & 7.01426\\
$\cdot$     & 7.0000 &   9.8995 &  14.0000 &  19.7990 &  28.0000 &\hspace{0.1in} 69 & 0.99497 &\hspace{0.1in} 68 & 7.01386\\
$\cdot$     & 8.0000 &  11.3137 &  16.0000 &  22.6274 &  32.0000 &\hspace{0.1in} 79 & 0.99498 &\hspace{0.1in} 77 & 7.01365\\
$\cdot$     & 9.0000 &  12.7279 &  18.0000 &  25.4558 &  36.0000 &\hspace{0.1in} 89 & 0.99498 &\hspace{0.1in} 87 & 7.01381\\
$\tilde{L}$ &10.0000 &  14.1421 &  20.0000 &  28.2843 &  40.0000 &\hspace{0.1in} 99 & 0.99499 &\hspace{0.1in} 97 & 7.01346\\
\hline\hline
\end{tabular}
\end{center}
\end{table}

The analogous calculation for the three-loop graph in Fig.~\ref{fig:basketball}
can also be done numerically.  The energy cutoff now limits the range
of indices for the four intermediate constituents.  Again there are
two time orderings, with contributions
\be
\widetilde{\Pi}_a(0)=\widetilde{\Pi}_b(0)
  =c\mu^2\int \frac{\prod_i^4 dp_{i-}\delta(\sum_i^4 p_{i-})}{(\prod_i^4 w_{p_{i-}})(\sum_i^4 w_{p_{i-}})}.
\ee
The numerical approximation suffers from the same limitations as in the one-loop
case, which is why the calculation of the spectrum at fixed box size 
in Sec.~\ref{sec:results} yields the divergent behavior in Fig.~\ref{fig:Evsc}.  
The approximation is
\be
\widetilde{\Pi}_a(0)=\widetilde{\Pi}_b(0)\simeq c\tilde{L}^2\sum_{n_1,n_2,n_3}^{\rm cutoff}
         \frac{1}{\left(\prod_i^4 \sqrt{n_i^2+c\tilde{L}^2}\right)\left(\sum_i^4\sqrt{n_i^2+c\tilde{L}^2}\right)},
\ee
with $n_4=-(n_1+n_2+n_3)$ and the cutoff specified by
\be
\frac{\mu}{\tilde{L}}\sum_i^4\frac{w_{n_i}-sn_i}{c}\leq E_{\rm max}.
\ee
One index is at its maximum $N$
when the other three are all equal to $-N/3$.  This gives $N$ as 
\be
N\leq \sqrt{c}\tilde{L}\sqrt{\left(\frac{\sqrt{c}E_{\rm max}}{2\mu}\right)^2
                             +4\left(\frac{2\mu}{\sqrt{c}E_{\rm max}}\right)^2-5}.
\ee
Values for this numerical approximation are tabulated in Table~\ref{tab:bubble}.
The convergence with respect to box size is quite rapid.

\subsection{Varied box size}  \label{sec:varied}

The three-loop vacuum bubble is embedded within the full eigenvalue
problem for the $\phi^4$ vacuum state.  Here we consider the
vacuum, in both the even and odd sectors, extrapolated in box size.
Because the box size is varied, we must consider the vacuum energy
density $E/2L$, rather than the (infinite) vacuum energy.  To do this,
we compute the lowest eigenvalue of $\frac{1}{2L}\Pplus$, with
$\Pplus=\Pplus^0+\Pplus^I$ specified by (\ref{eq:Pzero}) and
(\ref{eq:PplusI}).

The precise forms of the Hamiltonian terms are
\be
\frac{1}{2L}\Pplus^0=\frac{\mu^2}{2\pi}\sum_{n=-\infty}^\infty \frac{w_n-sn}{c\tilde{L}^2}a^\dagger_n a_n,
\ee
\bea
\frac{1}{2L}\Pplus^I=\frac{g\mu^2}{8\pi}\sum_{n_1\ldots n_4}\frac{1}{\sqrt{w_{n_1}\cdots w_{n_4}}}\left[
   \frac{1}{12}(a_{n_1}\cdots a_{n_4}+a^\dagger_{n_1}\cdots a^\dagger_{n_4})\delta_{n_1+\cdots+n_4,0}\right.&& \\
+\frac13(a^\dagger_{n_1}a_{n_2}a_{n_3}a_{n_4}+ a^\dagger_{n_2}a^\dagger_{n_3}a^\dagger_{n_4}a_{n_1})\delta_{n_1,n_2+n_3+n_4} &&
  \nonumber \\
\left.+\frac12 a^\dagger_{n_1}a^\dagger_{n_2}a_{n_3}a_{n_4}\delta_{n_1+n_2,n_3+n_4}\right], &&  \nonumber
\eea
with $w_n=\sqrt{n^2+c\tilde{L}^2}$.  In these we see that $c$ and $\tilde{L}$ appear only in the combination
$c\tilde{L}^2$. Thus, for any positive value of $c$, a rescaling of the box size makes the calculation
equivalent to an equal-time calculation at $c=1$, with a smaller box.  In the limit
of infinite box size, the calculation becomes completely independent of $c$, making 
the vacuum energy density independent of $c$, even as $c$ approaches the light-front limit
of zero.

For a numerical calculation, there is, of course, an energy cutoff $E_{\rm max}$.  For 
calculations with different values of $c$, this cutoff must be adjusted to make the basis
sizes equivalent.  In a Fock sector with $n$ constituents, the momentum indices lie
between $\pm N$ with 
\be \label{eq:N}
N=\sqrt{c}\tilde{L}\sqrt{\left(\frac{\sqrt{c}E_{\rm max}}{2\mu}\right)^2
   +\left(\frac{(n-1)^2-1}{4}\right)^2\left(\frac{2\mu}{\sqrt{c}E_{\rm max}}\right)^2
   -\frac{(n-1)^2+1}{2}}.
\ee
This is determined by giving one constituent $N$ units of momentum and each of the 
others $-N/(n-1)$ units, to minimize the total Fock-state energy 
\be
\frac{\mu}{\tilde{L}}\sum_n\frac{w_n-sn}{c}
 =\frac{\mu}{\tilde{L}}\left[\frac{\sqrt{N^2+c\tilde{L}^2}}{c}
                             +\frac{(n-1)}{c}\sqrt{\frac{N^2}{(n-1)^2}+c\tilde{L}^2}\right].
\ee
From (\ref{eq:N}) we see that, for fixed $N$, the energy cutoff must scale as 
$\mu/\sqrt{c}$, which is not surprising, given that
zero-mode energies diverge as $\mu/\sqrt{c}$.

For fixed $\sqrt{c}E_{\rm max}$, the size of the calculation saturates, with $N$ forced
to be zero in the highest Fock sectors.  Setting $N=0$ and $n=K$ in (\ref{eq:N}),
we find that this occurs at a maximum number $K$ of constituents, given by
\be
K=1\mp2\pm\sqrt{8-2(1-2\sqrt{c}E_{\rm max}/\mu)}.
\ee
For $E_{\rm max}=20\mu$ and $c=1$, this
limits $K$ to 8 (9) in the even (odd) sector.  Beyond these values of $K$,
only zero modes contribute.

With calculations at arbitrary $c$ now shown to be completely equivalent to 
equal-time calculations, the light-front limit must then reproduce the results
found for equal-time quantization~\cite{RychkovVitale,Rychkov}.  In particular, 
they must agree on the value of the critical coupling.

\section{Summary}
\label{sec:summary}

By considering the coordinate interpolation (\ref{eq:coordinates}), we
have been able to study the approach to light-front quantization
($c=0$) from equal-time quantization ($c=1$) in two-dimensional
$\phi^4$ theory. A numerical calculation of the spectrum, for arbitrary $c>0$
and fixed box size, provides results consistent with those of equal-time 
quantization~\cite{RychkovVitale,Rychkov}, as shown in Fig.~\ref{fig:Diffvsgc}.  
However, the spectrum is found to diverge as $c$ approaches the light-front 
limit of zero, if the box size is held fixed; see Fig.~\ref{fig:Evsc}.  
Although this might be taken as an indication
that the light-front limit is not smooth, we have shown that the 
divergence is instead spurious and caused by a poor numerical representation
of vacuum-bubble contributions.  The spurious divergence is due
to the fixed box size, which prevents the momentum-space grid from
being fine enough to sample the vacuum-bubble integrals accurately.

Our nonperturbative analysis of vacuum-bubble contributions replicates the
perturbative analysis of Collins~\cite{Collins}.  The contributions
of the two time-ordered graphs of Fig.~\ref{fig:1looptimeordered} are shown
to be different, depending on the order of the limits $c\rightarrow0$ and
$p^2\rightarrow0$; the sum, however, is invariant.  The numerical 
approximation to these contributions is then shown to be very
sensitive to box size.  In particular, the $c\rightarrow0$ limit
requires that the continuum limit in momentum space (infinite
box size) must be taken as $c\rightarrow0$.  The DLCQ~\cite{PauliBrodsky}
formulation of light-front quantization, with its finite box, will
require a different approach for the inclusion of vacuum-bubble contributions.

The vacuum energy density is shown in Sec.~\ref{sec:varied} to be
independent of $c$ in the continuum limit of infinite box size.  This
makes a calculation for any $c>0$ explicitly equivalent to an
equal-time calculation at $c=1$.  Assuming that the $c\rightarrow0$ limit
is smooth, the light-front spectrum must be nonperturbatively equivalent
to the equal-time spectrum, provided vacuum-bubble contributions are taken
into account properly.  Thus, the apparent disagreement over the
value of the critical coupling, resolved by noting differences
in mass renormalization~\cite{BCH}, is actually a sign that
vacuum effects are not included properly.  How to include
vacuum bubbles in a native, nonperturbative light-front calculation
remains an open question.

\acknowledgments
This work was supported in part by 
the Minnesota Supercomputing Institute through
grants of computing time and benefited from 
participation in the workshop on
Hamiltonian methods in strongly coupled quantum field theory
supported by the Simons Collaboration on the Nonperturbative Bootstrap.

\appendix

\section{Nonperturbative one-loop bubble from $\phi^3$ theory}  \label{sec:phi3theory}

Contributions to the energy that correspond to
the time-ordered loops in Fig.~\ref{fig:1looptimeordered}
arise naturally in a truncated nonperturbative calculation,
which we show here.  The interaction Hamiltonian for $\phi^3$ theory
is 
\be
\Pplus^I=\int_{-\infty}^\infty dx^-\frac{\lambda}{3!}:\phi^3:.
\ee
Substitution of the mode expansion (\ref{eq:continuousmodeexp}) 
and integration over $x^-$ yields
\bea
\Pplus^I&=&\frac{\lambda}{4\sqrt{4\pi}}\int\frac{dp_{1-}dp_{2-}dp_{3-}}{\sqrt{w_{p_1}w_{p_2}w_{p_3}}}
\left\{a^\dagger(p_{1-})a(p_{2-})a(p_{3-})\delta(p_{1-}-p_{2-}-p_{3-})\right. \\
  &&     +a^\dagger(p_{1-})a^\dagger(p_{2-})a(p_{3-})\delta(p_{1-}+p_{2-}-p_{3-}) \nonumber \\
&&\left. +\frac13[a(p_{1-})a(p_{2-})a(p_{3-}
    )+a^\dagger(p_{1-})a^\dagger(p_{2-})a^\dagger(p_{3-})]\delta(p_{1-}+p_{2-}+p_{3-})\right\}.
       \nonumber
\eea
The eigenstate is expanded in a Fock basis as
\bea
\lefteqn{|\psi(P_-)\rangle=\psi_1a^\dagger(P_-)|0\rangle
+\int dp_{1-}dp_{2-}\delta(P_--p_{1-}-p_{2-})\psi_2(p_{1-},p_{2-})\frac{1}{\sqrt{2}}a^\dagger(p_{1-})a^\dagger(p_{2-})|0\rangle}&& 
  \nonumber \\
&&+\int dp_{1-}dp_{2-}dp_{3-}\delta(P_--p_{1-}-p_{2-}-p_{3-})\psi_3(p_{1-},p_{2-},p_{3-})
     \frac{1}{\sqrt{6}}a^\dagger(p_{1-})a^\dagger(p_{2-})a^\dagger(p_{3-})|0\rangle \nonumber \\
&&+\int dp_{1-}dp_{2-}dp_{3-}dp_{4-}\delta(P_--p_{1-}-p_{2-}-p_{3-}-p_{4-})\psi_4(p_{1-},p_{2-},p_{3-},p_{4-})  \\
  && \hspace{0.5in} \times   \frac{1}{\sqrt{24}}a^\dagger(p_{1-})a^\dagger(p_{2-})a^\dagger(p_{3-})a^\dagger(p_{4-})|0\rangle,  \nonumber
\eea
with truncation at four constituents.  The action of the free part of the Hamiltonian is
\bea
\lefteqn{\Pplus^0|\psi(P_-)\rangle=\frac{w_P-sP_-}{c}\psi_1a^\dagger(P_-)|0\rangle}&& \\
&&+\int dp_{1-}dp_{2-}\delta(P_--p_{1-}-p_{2-})\sum_i^2\left[\frac{w_{p_i}-sp_{i-}}{c}\right]
             \psi_2(p_{1-},p_{2-})\frac{1}{\sqrt{2}}a^\dagger(p_{1-})a^\dagger(p_{2-})|0\rangle \nonumber \\
&&+\int dp_{1-}dp_{2-}dp_{3-}\delta(P_--p_{1-}-p_{2-}-p_{3-}) \nonumber \\
   && \hspace{0.5in} \times  \sum_i^3\left[\frac{w_{p_i}-sp_{i-}}{c}\right]\psi_3(p_{1-},p_{2-},p_{3-})
     \frac{1}{\sqrt{6}}a^\dagger(p_{1-})a^\dagger(p_{2-})a^\dagger(p_{3-})|0\rangle \nonumber \\
&&+\int dp_{1-}dp_{2-}dp_{3-}dp_{4-}\delta(P_--p_{1-}-p_{2-}-p_{3-}-p_{4-})  \nonumber \\
  && \hspace{0.5in} \times  \sum_i^4\left[\frac{w_{p_i}-sp_{i-}}{c}\right]\psi_4(p_{1-},p_{2-},p_{3-},p_{4-}) 
        \frac{1}{\sqrt{24}}a^\dagger(p_{1-})a^\dagger(p_{2-})a^\dagger(p_{3-})a^\dagger(p_{4-})|0\rangle  \nonumber
\eea
and that of the interaction part, truncated at four constituents, is
\bea
\Pplus^I|\psi(P_-)\rangle&=&\frac{\lambda}{4\sqrt{4\pi}}\left\{
      \int \frac{dp_{1-}dp_{2-}}{\sqrt{w_{p_1}w_{p_2}w_{P}}}
       \delta(p_{1-}+p_{2-}-P_-)\psi_1 a^\dagger(p_{1-})a^\dagger(p_{2-})|0\rangle \right. \\
      && +\frac13\int \frac{dp_{1-}dp_{2-}dp_{3-}}{\sqrt{w_{p_1}w_{p_2}w_{p_3}}}
       \delta(p_{1-}+p_{2-}+p_{3-})\psi_1 a^\dagger(p_{1-})a^\dagger(p_{2-})a^\dagger(p_{3-})a^\dagger(P_-)|0\rangle 
           \nonumber \\
      && +\sqrt{2}\int \frac{dp_{1-}dp_{2-}}{\sqrt{w_{p_1}w_{p_2}w_{P}}}
       \delta(p_{1-}+p_{2-}-P_-)\psi_2(p_{1-},p_{2-})a^\dagger(P_-)|0\rangle\nonumber \\
      && +\sqrt{2}\int \frac{dp_{1-}dp_{2-}dp_{3-}}{\sqrt{w_{p_1}w_{p_2}w_{p_1+p_2}}}
       \delta(p_{1-}+p_{2-}+p_{3-}-P_-)\nonumber \\
       && \hspace{0.5in} \times\psi_2(p_{1-}+p_{2-},p_{3-})
           a^\dagger(p_{1-})a^\dagger(p_{2-})a^\dagger(p_{3-})|0\rangle \nonumber \\
      && +\sqrt{6}\int \frac{dp_{1-}dp_{2-}dp_{3-}}{\sqrt{w_{p_1}w_{p_2}w_{p_1-p_2}}}
       \delta(p_{1-}+p_{3-}-P_-)\nonumber \\
       && \hspace{0.5in} \times\psi_3(p_{1-}-p_{2-},p_{2-},p_{3-})
           a^\dagger(p_{1-})a^\dagger(p_{3-})|0\rangle \nonumber \\
      && +\sqrt{\frac32}\int \frac{dp_{1-}dp_{2-}dp_{3-}dp_{4-}}{\sqrt{w_{p_1}w_{p_2}w_{p_1+p_2}}}
       \delta(p_{1-}+p_{2-}+p_{3-}+p_{4-}-P_-) \nonumber \\
       && \hspace{0.5in} \times\psi_3(p_{1-}+p_{2-},p_{3-},p_{4-})
           a^\dagger(p_{1-})a^\dagger(p_{2-})a^\dagger(p_{3-})a^\dagger(p_{4-})|0\rangle \nonumber \\
      &&  +\sqrt{\frac83}\int \frac{dp_{1-}dp_{2-}dp_{3-}}{\sqrt{w_{p_1}w_{p_2}w_{p_3}}}\delta(p_{1-}+p_{2-}+p_{3-})
         \psi_4(p_{1-},p_{2-},p_{3-},P_-)a^\dagger(P_-)|0\rangle  \nonumber \\
     &&   +\sqrt{6}\int \frac{dp_{1-}dp_{2-}dp_{3-}dp'_{1-}}{\sqrt{w_{p_1}w_{p'_1}w_{p_1-p'_1}}}
       \delta(p_{1-}+p_{2-}+p_{3-}-P_-)  \nonumber \\
       && \hspace{0.5in} \left.\times \psi_4(p'_{1-},p_{1-}-p'_{1-},p_{2-},p_{3-})
           a^\dagger(p_{1-})a^\dagger(p_{2-})a^\dagger(p_{3-})|0\rangle   \right\}.  \nonumber
\eea

\begin{figure}[ht]
\vspace{0.2in}
\centerline{\includegraphics[width=7cm]{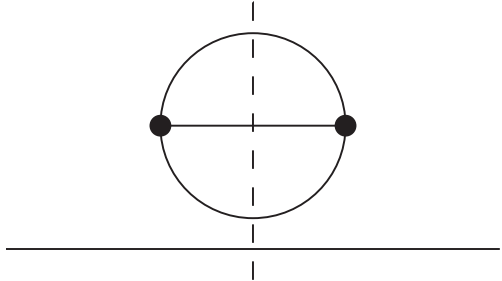}}
\caption{\label{fig:2loopbubble}
Disconnected two-loop bubble in $\phi^3$ theory.}
\end{figure}

From these contributions, we can construct the eigenvalue problem $\Pplus|\psi(P_-)\rangle=P_+|\psi(p_-)\rangle$
as projections onto different Fock sectors.  These projections are
\bea  \label{eq:psi1}
\lefteqn{\frac{w_P-sP_-}{c}\psi_1
   +\sqrt{2}\frac{\lambda}{4\sqrt{4\pi}}\int\frac{dp_{1-}dp_{2-}}{\sqrt{w_Pw_{p_1}w_{p_2}}}
                  \delta(P_--p_{1-}-p_{2-})\psi_2(p_{1-},p_{2-})} && \\
 &&  +\sqrt{\frac83}\frac{\lambda}{4\sqrt{4\pi}}\int\frac{dp_{1-}dp_{2-}dp_{3-}}{\sqrt{w_{p_1}w_{p_2}w_{p_3}}}
                  \delta(p_{1-}+p_{2-}+p_{3-})\psi_4(p_{1-},p_{2-},p_{3-},P_-)
   =P_+\psi_1, \nonumber
\eea
\bea  \label{eq:psi2}
\lefteqn{\sum_i^2\frac{w_{p_i}-sp_{i-}}{c}\psi_2(p_{1-},p_{2-})
  +\frac{\lambda}{4\sqrt{4\pi}}\frac{\sqrt{2}\psi_1}{\sqrt{w_{p_1}w_{p_2}w_{P}}}} && \\
  && +\sqrt{12}\frac{\lambda}{4\sqrt{4\pi}}\int \frac{dp'_{2-}}{\sqrt{w_{p_1}w_{p'_2}w_{p_1-p'_2}}}
                          \psi_3(p_{1-}-p'_{2-},p'_{2-},p_{2-})=P_+\psi_2,  \nonumber
\eea
\bea  \label{eq:psi3}
\lefteqn{\sum_i^3\frac{w_{p_i}-sp_{i-}}{c}\psi_3(p_{1-},p_{2-},p_{3-})
    +\frac{\lambda}{4\sqrt{4\pi}}\frac{\sqrt{12}\psi_2(p_{1-}+p_{2-},p_{3-})}{\sqrt{w_{p_1}w_{p_2}w_{p_1+p_2}}}} && \\
  && +6\frac{\lambda}{4\sqrt{4\pi}}\int \frac{dp'_{1-}}{\sqrt{w_{p_1}w_{p'_1}w_{p_1-p'_1}}}
        \psi_4(p'_{1-},p_{1-}-p'_{1-},p_{2-},p_{3-})=P_+\psi_3,  \nonumber
\eea
\bea   \label{eq:psi4}
\lefteqn{\sum_i^4\frac{w_{p_i}-sp_{i-}}{c}\psi_4(p_{1-},p_{2-},p_{3-},p_{4-})
    +\frac{\lambda}{4\sqrt{4\pi}}\frac{6\psi_3(p_{1-}+p_{2-},p_{3-},p_{4-}}{\sqrt{w_{p_1}w_{p_2}w_{p_1+p_2}}}} && \\
    && +\frac{\lambda}{4\sqrt{4\pi}}\frac{\psi_1}{\sqrt{6}}
      \left[\frac{\delta(p_{1-}+p_{2-}+p_{3-})}{\sqrt{w_{p_1}w_{p_2}w_{p_3}}}
            +\frac{\delta(p_{1-}+p_{3-}+p_{4-})}{\sqrt{w_{p_1}w_{p_3}w_{p_4}}} \right. \nonumber \\
    && \left. \hspace{1in}        +\frac{\delta(p_{1-}+p_{2-}+p_{4-})}{\sqrt{w_{p_1}w_{p_2}w_{p_4}}}
            +\frac{\delta(p_{2-}+p_{3-}+p_{4-})}{\sqrt{w_{p_2}w_{p_3}w_{p_4}}}\right]
    =P_+\psi_4.  \nonumber
\eea
We next invert the equations (\ref{eq:psi2}) and (\ref{eq:psi4}) for $\psi_2$ and $\psi_4$, 
neglecting the higher-order corrections that come from the $\psi_3$-contributions determined
by (\ref{eq:psi3}), and substitute into the equation (\ref{eq:psi1}) for $\psi_1$.  This 
leaves a single implicit equation for $P_+$, with $\psi_1$ as a common factor, which we remove.
After some simplications that combine algebraic factors and that integrate over all but
one delta function, we obtain
\bea
P_+&=&\frac{w_P-sP_-}{c}
  + \frac{\lambda^2}{32\pi}\int\frac{dp_{1-}dp_{2-}}{w_{p_1}w_{p_2}w_P}
  \frac{\delta(P_--p_{1-}-p_{2-})}{P_+-\sum_i^2\frac{w_{p_i}-2p_{i-}}{c}} \\
  &&   + \frac{\lambda^2}{32\pi}\int\frac{dp_{1-}dp_{2-}}{w_{p_1}w_{p_2}w_P}
  \frac{\delta(P_-+p_{1-}+p_{2-})}{P_+-2\frac{w_P-sP_-}{c}-\sum_i^2\frac{w_{p_i}-2p_{i-}}{c}} \nonumber \\
  &&   + \frac13\delta(0)\frac{\lambda^2}{32\pi}\int\frac{dp_{1-}dp_{2-}dp_{3-}}{w_{p_1}w_{p_2}w_{p_3}}
  \frac{\delta(p_{1-}+p_{2-}+p_{3-})}{P_+-\sum_i^3\frac{w_{p_i}-2p_{i-}}{c}-\frac{w_P-sP_-}{c}}. \nonumber
\eea
The second and third terms on the right correspond to the time-ordered graphs
in Fig.~\ref{fig:1loopbubble}.  The last term is a disconnected two-loop bubble, which injects
an infinite constant to be subtracted; it corresponds to the graph in Fig.~\ref{fig:2loopbubble}.
The result quoted in (\ref{eq:P+phi3}) is just this equation after subtraction of the 
two-loop bubble.


\end{document}